\documentclass[journal=acsnano]{achemso}
\setkeys{acs}{articletitle=true}
\usepackage{lineno,hyperref,amssymb,caption,subcaption,graphicx,float}

\vspace{-1em}
\title{Very High Thermoelectric Power Factor Near Magic Angle in Twisted Bilayer Graphene} \vspace{-1em}
\author{Adithya Kommini}
\author{Zlatan Aksamija}\vspace{-1em}
\email{zlatana@engin.umass.edu}\vspace{-1em}
\affiliation{University of Massachusetts Amherst, Amherst, MA 01003}
	
\begin{document}
\vspace{-1em}
\begin{abstract}

Recent research on twisted bilayer graphene (TBG) uncovered that its twist-angle-dependent electronic structure leads to a host of unique properties, such as superconductivity, correlated insulating states, and magnetism. The flat bands that emerge at low twist angles in TBG result in sharp features in the electronic density of states (DOS), affecting transport. Here we show that they lead to superior and tuneable thermoelectric (TE) performance. Combining an iterative Boltzmann transport equation solver and electronic structure from an exact continuum model, we calculate thermoelectric transport properties of TBG at different twist angles, carrier densities, and temperatures. Our simulations show the room-temperature TE power factor (PF) in TBG reaches 40 mWm$^{-1}$K$^{-2}$, significantly higher than single-layer graphene (SLG) and among the highest reported to date. The peak PF is observed near the magic angle, at a twist angle of $\approx$1.3$^\circ$, and near complete band filling. We elucidate that its dependence is driven by the band gap between lowest and second subband, which improves Seebeck but decreases with twist angle, and Fermi velocity, which dictates conductivity and rises with twist angle. We observed a further increase in the PF of TBG with decreasing temperature. The strong TE performance, along with the ability to fine-tune transport using twist angle, makes TBG an interesting candidate for future research and applications in energy conversion and thermal sensing and management.
\vspace{-1em}

\end{abstract}
\maketitle




Ever-increasing demand for sustainable energy resources amplifies the need to improve the efficiency of existing power generation techniques and develop novel approaches across a broad swath of size scales. Waste-heat recovery through solid-state thermoelectric (TE) devices is one such approach that can contribute to robust small-scale energy scavenging and thermal management. Their widespread adoption may require new materials with high TE performance, typically measured by the dimensionless figure-of-merit $ZT$ \cite{Goldsmid1964}. ZT is proportional to the power factor (PF), given by the product $S^2 \sigma$ of electrical conductivity $\sigma$ and the Seebeck coefficient $S$ squared. There is a well-established trade-off between $\sigma$, which increases with carrier concentration, and $S$, which decreases with it. Band-engineering techniques \cite{Pei2012a,SnyderNMAT08,YanSciRep2017} have been pursued to decouple this dependency and improve the PF by introducing sharp features in the electronic density-of-state (DOS), whose large slope benefits $S$. A complementary approach is by employing low-dimensional materials, such as graphene, that naturally possess a suitable DOS. Single-layer graphene (SLG) on hexagonal boron nitride (hBN) substrates has achieved the highest known room-temperature PF of 36 mW m$^{-1}$ K$^2$ \cite{DuanPNAS2016}. Beyond power generation applications, high-PF materials like graphene can enhance active cooling, which is vital for thermal management in electronic devices \cite{ZebarjadiAPL15}. Furthermore, graphene thermocouples exhibit the highest reported sensitivity \cite{PascalAFM2020}, which can be used in reliable temperature sensors. 

Recent studies showed that absence of Fermi velocity at certain magic angles in twisted bilayer graphene (TBG) results in unconventional superconductivity \cite{Cao2018,Cao2018a} at low temperatures. The theoretically predicted flat bands near magic angles \cite{BistritzerPNAS11,MacDonald2011} contribute to superconducting states depending on the carrier density and twist angle. These remarkably flat bands with strongly correlated states also exhibit Mott-like insulator behavior at half-filling of the moir\'e subbands. The flat bands indicate van Hove singularities in the carrier DOS near the Fermi level~\cite{LiNatPhys10}, which, if coupled with good carrier conductivity, is highly desirable for thermoelectric applications. This combination of features in the DOS, in addition to the secondary bandgap between the lowest and second subbands in TBG, creates a narrow transport distribution function (TDF) \cite{KomminiJCE16}. Then states bunched around sharp features in the DOS translate into improved Seebeck coefficient and maximize the PF\cite{Mahan1989,KomminiJOPM2019}. The highly tunable band structure of TBG offers a new lever of control and a novel avenue to further decouple Seebeck and conductivity to achieve unprecedented PFs. However, little is yet known about the TE performance of TBG and its dependence on twist angle except at very low temperatures \cite{Kubakaddi2020}. 

Here we study the impact of the unique band structure and sharp features in the DOS on carrier transport and thermoelectric properties of TBG. We numerically solve the Boltzmann transport equation (BTE) using band structure obtained from a recently published exact continuum model coupled with a full-band variant of Rode\textquoteright s iterative method to model the carrier transport in TBG. Our model shows TBG has an exceptionally high PF near, but not coinciding with, the magic twist angle. We explain this behavior via the trade-off between bands flattening on the one hand, which benefits the Seebeck, and Fermi velocity dropping, which lowers conductivity. The peak room-temperature PF of 1 Wm$^{-1}$K$^{-2}$ is the highest ever reported and more than an order of magnitude higher than in SLG. Compared to SLG, the PF in TBG is driven by a more favorable trade-off between conductivity and Seebeck, which reaches up to 150 $\mu$VK$^{-1}$. We observe further increases in the PF when the temperature is lowered to 200 and 100 K, consistent with the presence of a small band gap. 

We have implemented a self-consistent Jacobi iterative scheme to solve the linearized BTE and include the full band structure as input \cite{KomminiJPCM2018,KomminiPRApp20}. Our approach to solving the BTE is based on Rode\textquoteright s method \cite{Rode1971} to incorporate inelastic scattering, particularly the in-scattering terms in the collision integral. This is important in a Dirac material like graphene where transport is predominantly bipolar and significantly affected by inelastic scattering \cite{ZuevPRL09}. The carrier probability can be written as a sum of the equilibrium component (determined by Fermi-Dirac statistics, ${f}_{0}(k)$) and a small perturbation (${g}{\left(k\right)}$) in the direction of the applied field, ${f(k)}={f}_{0}(k)+g(k)$. Using the standard Boltzmann form of the collision operator, the perturbation to carrier distribution is calculated by iterating
\begin{equation}\label{Rode_b}
{g}_{i+1}(k)=\left({I_i(k)+eF\nabla_k {f}_{0}-v(k)\cdot\nabla_r{f}_{0}}\right) /{{S}_{o}(k)},
\end{equation}
\noindent where the in-scattering integral is
\begin{equation}\label{I_k}
I_i(k)={\Lambda} \int \left[N(\omega_{ph}) - f(k) + \frac{1}{2} \mp \frac{1}{2}\right] {g}_{i}\left({k}^{\prime}\right) \delta(E(k) \pm \hbar \omega_{ph}-E(k^{\prime})) dk',
\end{equation}
\noindent $S_o$ is the total scattering rate (inverse of relaxation time) and $F$ is the applied electric field. The term $eF\nabla_k {f}_{0}$ can be expanded to get $(e/\hbar) F\cdot v(k)\partial f_0/\partial E$ with $v(k)=1/\hbar \nabla_k E(k)$ being the carrier group velocity while the $\partial f_0/\partial E$ is the Fermi window that determines the range of energies contributing to transport.
\begin{figure*}[ht!]
	\centering
	\includegraphics[width=1\columnwidth]{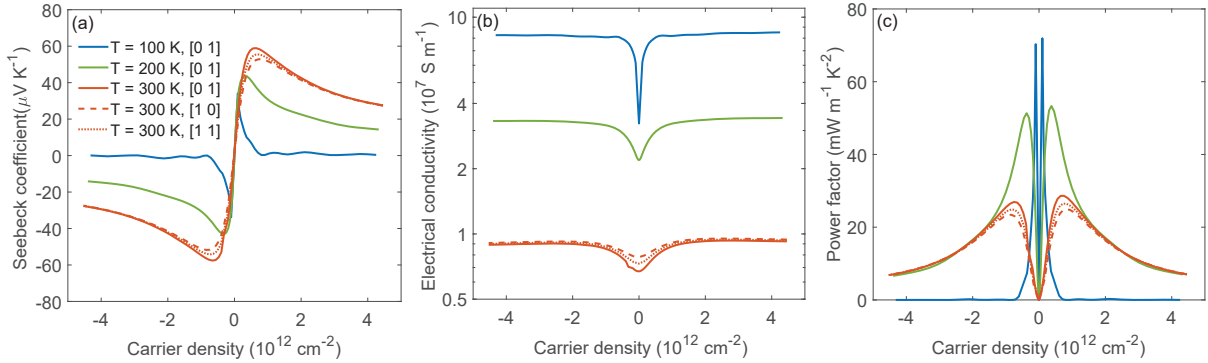}
	\caption{Thermoelectric properties of SLG a) Seebeck coefficient, b) electrical conductivity, and c) PF calculated at different temperatures. The peak in S is observed when applied electric field in [0 1] direction at T = 300 K. This translates to higher power factor in [0 1] direction. S$_{eff}$ decreases with decreasing temperature while $\sigma_{eff}$ increases.}
	\label{fig1}
\end{figure*}

In strongly correlated systems or "strange metals", resistivity scales linearly with temperature, similar to what was observed in TBG \cite{Marel2003,Cao2020}. In these materials, quantum fluctuations are proposed to be the dominant scattering mechanism that results in T-linear behavior \cite{Emery1995,Bruin2013}. However, the T-linear resistivity is observed at a broad range of twist angles in TBG, independent of the correlated phases at magic angles \cite{Polshyn2019}. This indicates a unified scattering mechanism controls carrier transport. Phonons have been identified as that mechanism \cite{Wu2018} and inelastic phonon scattering was shown to dominate carrier transport above the superconducting critical  temperature $T_c$ \cite{Yudhistira2019}. Hence this study focuses on simulating phonon-limited carrier transport at and above 100 K, well in excess of the Bloch-Gr\"uneisen temperature \cite{HwangPRB08} and in the range of typical energy and sensing applications. Lower temperatures also result in hydrodynamic transport \cite{Zarenia2020} and an enhanced contribution from phonon drag \cite{Kubakaddi2010}, which are beyond the scope of our model. The scattering processes included here are acoustic phonon scattering ($\Gamma_{ac}$)
\begin{equation}\label{Gamma_ac}
\Gamma_{ac}(E) = \frac{2\pi D_{ac}^2 k_B T}{\hbar \rho v_{ac}^2} \int \delta(E-E(k')) dk' = \frac{2\pi D_{ac}^2 k_B T}{\hbar \rho v_{ac}^2} D(E),\\
\end{equation}
where $D(E)$ is the DOS, and inelastic optical phonon scattering ($\Gamma_{op}$)
\begin{equation}\label{Gamma_op}
\Gamma_{op}(k) = \frac{\pi D_{op}^2 }{\rho \omega_{op}} \int \left[N(\omega_{op}) - f(k')+ \frac{1}{2} \mp \frac{1}{2}\right] \delta(E(k)\pm \hbar\omega_{op}-E(k')) dk' ,
\end{equation}
where the top sign represents absorption processes and bottom emission processes. Here, $D_{ac}$ is the acoustic deformation potential, $D_{op}$ is the optical deformation potential, $\rho$ is the mass density, $v_{ac}$ is the sound velocity of acoustic phonon modes, $\omega_{op}$ is the optical phonon frequency, $N(\omega_{op})$ is phonon distribution function and $f(k')$ is the carrier probability distribution of the final state. In Eq. \ref{Gamma_op}, the constants in front of the integral together make up the scattering-dependent pre-factor $\Lambda$ used in Eq. \ref{I_k}~\cite{KomminiJPCM2018}. 

The Brillouin zone integrals are evaluated numerically using a 2D version of Spherical Averaging (SAVE) method \cite{CorreaNT17}. In SAVE, we linearly extrapolate the band structure inside a small circle of radius $R_s$ around each point $k^\prime$ in the discretization to compute the length of the constant-energy contour, then sum all the contributions with a weight of $\Lambda(k^{\prime})g_i(k^\prime)$. Once the Jacobi iteration reaches convergence, as measured by the relative error between successive iterations, the converged $g(k)$ is used to calculate transport distribution function (TDF) as 
\begin{equation}\label{TDF}
\sigma(E) = \frac{1}{\Omega (2\pi)^2 F} \int v(k) g(k) \delta(E-E(k)) d^2k 
\end{equation}
\noindent where $\Omega$ is the volume of the first Brillouin zone. The $S$ and $\sigma$ of each carrier type are calculated from the TDF as
\begin{equation}
\sigma = \int \sigma(E) dE, \\
\end{equation}
\begin{equation}
S = -\frac{1}{eT} \frac{\int \sigma(E)(E-E_f) dE}{\int \sigma(E) dE}
\end{equation}
\noindent with $E_f$ being Fermi energy level and $T$ the temperature of the material. Due to the bipolar nature of the transport in TBG, the effective Seebeck coefficient ($S_{eff}$) is calculated by combining the electron ($S_n$) and hole Seebeck coefficient ($S_p$) over their respective conductivities ($\sigma_n$ and $\sigma_p$), $S_{eff}$ = ($S_n  \sigma_n + S_p  \sigma_p$)/($\sigma_n+\sigma_p$). The total carrier conductivity is the sum of electron and hole conductivity, $\sigma_{eff}$ = $\sigma_n+\sigma_p$.

In order to validate our model, we first calculated TE properties of SLG, using electronic band structure obtained from first principles in our previous work \cite{Yasaei2015}. Scattering rates are calculated using Eq. \ref{Gamma_ac} and Eq. \ref{Gamma_op} with D$_{ac}$ = 3 eV, D$_{op}^\Gamma$ = 100 eV nm$^{-1}$ at $\Gamma$-point, and D$_{op}^K$ = 200 eV nm$^{-1}$ at K-point. The deformation potentials are taken from \cite{Borysenko2010a} and adjusted to match the Seebeck coefficient~\cite{KanahashiNPJ2019} and maximum power factor~\cite{ZuevPRL09} from the literature. The inelastic optical phonon processes are calculated at optical phonon energies of $\omega_{op}^\Gamma$ = 1589 cm$^{-1}$ = 0.197 eV at $\Gamma$-point and $\omega_{op}^K$ = 1197 cm$^{-1}$ = 0.148 eV at K-point taken from Ref.~\cite{Cocemasov2013}. Fig. \ref{fig1}(a) shows the $S_{eff}$ calculated for SLG at different temperatures by varying the carrier densities using Rode\textquoteright s method. Our $S_{eff}$ and  $\sigma_{eff}$ (Fig.\ref{fig1}) for SLG are in agreement with measurements on hBN-encapsulated samples~\cite{ZuevPRL09,DuanPNAS16}. However, scattering from charged impurities in the environment can impact transport, lowering conductivity but raising the Seebeck \cite{HwangPRB09}. We neglect impurities here as their location and concentration varies from sample to sample depending on preparation conditions. A peak PF of 28.6 mW m$^{-1}$ K$^2$ is observed in SLG for electrons and 30 mW m$^{-1}$ K$^2$ for holes at 300 K in Fig.\ref{fig1}(c) when electric field is applied in [0 1] direction, in good agreement with measurements \cite{KanahashiNPJ2019,DuanPNAS16}, confirming the need to include the contribution from inelastic processes~\cite{Ghahari2016}. 


\begin{figure*}[ht!]
	\centering
	\includegraphics[width=1\columnwidth]{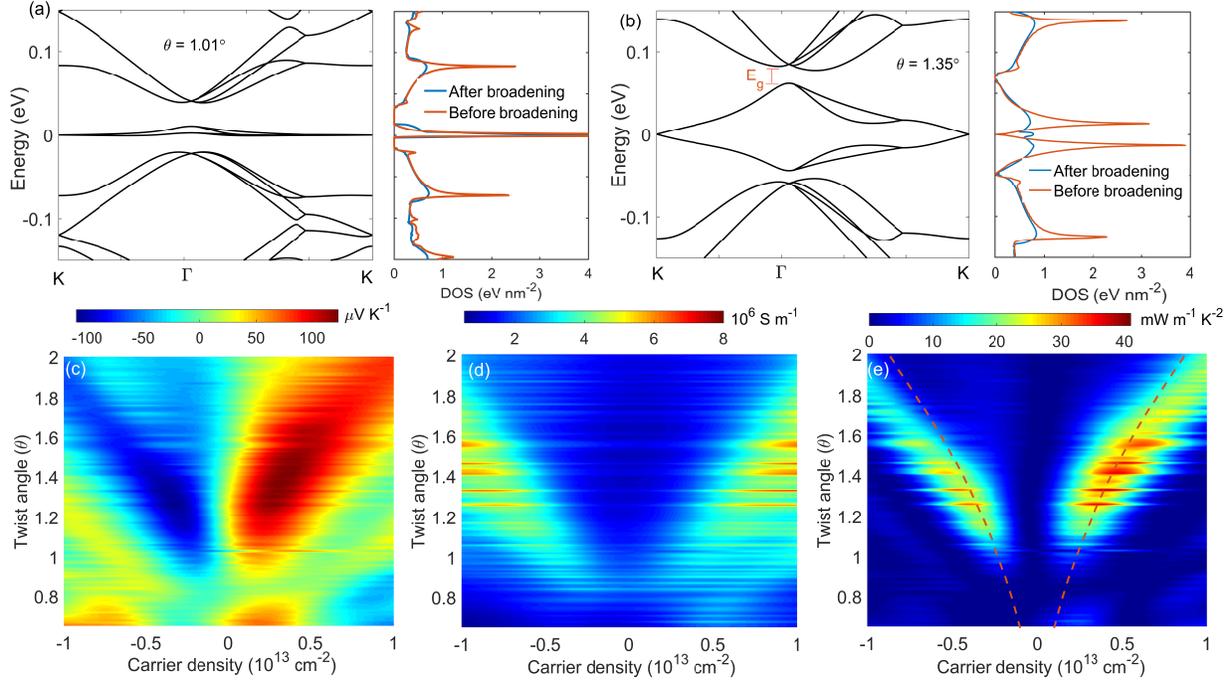}
	\caption{Band structure and density of states (DOS) of TBG at different twist angles a) 1.01$^\circ$ and b) 1.35$^\circ$. Room temperature c) Seebeck coefficient($S$), d) electrical conductivity ($\sigma$), and e) PF of TBG at different twist angles. The peak in power factor is observed at a twist angle of 1.33$^\circ$. At this twist angle, we observe a maximum in S, which results in the highest power factor. The dotted lines represent $\pm n_s$, the carrier density required to completely fill the lowest moir\'e subband.}
	\label{Fig2}
\end{figure*}


Next, our model was used for TBG to calculate the TE properties at different twist angles. The twist-angle-resolved band structure of TBG used in this study is obtained from an openly available exact continuum model developed in Ref.~\cite{CarrPhyRevRes2019,Fang2019}. The exact continuum model is based on $k\cdot p$ perturbation theory, combining the computational efficiency of tight-binding Hamiltonians with the twist-angle control offered by continuum models. In Fig.~\ref{Fig2}(a) and Fig.~\ref{Fig2}(b), we plot two representative examples of electronic band structures used in this study at twist angles $\theta$ = 1.01$^\circ$ and $\theta$ = 1.35$^\circ$, respectively, showing the flat bands near magic angle; the low energy bands open up with increasing twist angle (Fig.~\ref{Fig2}(b)) and eventually recover the Dirac cone at the K-point. The secondary band gap (E$_g$ shown in \ref{Fig2}(b)) between the first and second subbands of electrons and holes influences carrier transport, especially at half-filling where the material exhibits correlated insulator behavior.\cite{CaoNature18}

Due to the presence of flat bands and gaps, the DOS of TBG has sharp features that are affected by interactions with the lattice at temperatures of interest here. Collision broadening \cite{Kim1987,Reggiani1987} from carrier scattering with phonons is implemented in this study by relating the scattering rate to the imaginary part of self energy $\Gamma(k)=-(2/\hbar)\Im \Sigma(k)$ and replacing the energy-conserving $\delta$ in Eqs.~\ref{I_k}-\ref{Gamma_op} with a Lorentzian spectral function $(1/\pi) \Gamma(k)/[(E(k)\pm \hbar\omega_{ph}-E(k'))^2 + \Gamma(k)^2]$~\cite{AksamijaJAP09,Pletikosic2012}. The DOS calculated after applying collision broadening is shown in Fig. \ref{Fig2}(a) and Fig. \ref{Fig2}(b) and compared with the unbroadened DOS to show that broadening partially smooths out some of the sharp features in the DOS. Acoustic phonon scattering in TBG is calculated with an acoustic deformation potential of D$_{ac}$ = 9 eV. This is in line with literature values that suggest D$_{ac}$ of TBG is 3 times higher than the value observed in SLG~\cite{Borysenko2011,Polshyn2019,KoshinoPRB20}. D$_{op}^\Gamma$ = 10 eV nm$^{-1}$ with optical phonon energy of 0.150 meV (1213 cm$^{-1}$) at $\Gamma$-point \cite{Cocemasov2013} and D$_{op}^K$ = 20 eV nm$^{-1}$ ~\cite{Borysenko2011} with optical phonon energies of 0.148 meV (1197 cm$^{-1}$) at K-point are used to calculate the inelastic phonon scattering rates.

\begin{figure*}[ht!]
	\centering {\includegraphics[width=1\columnwidth]{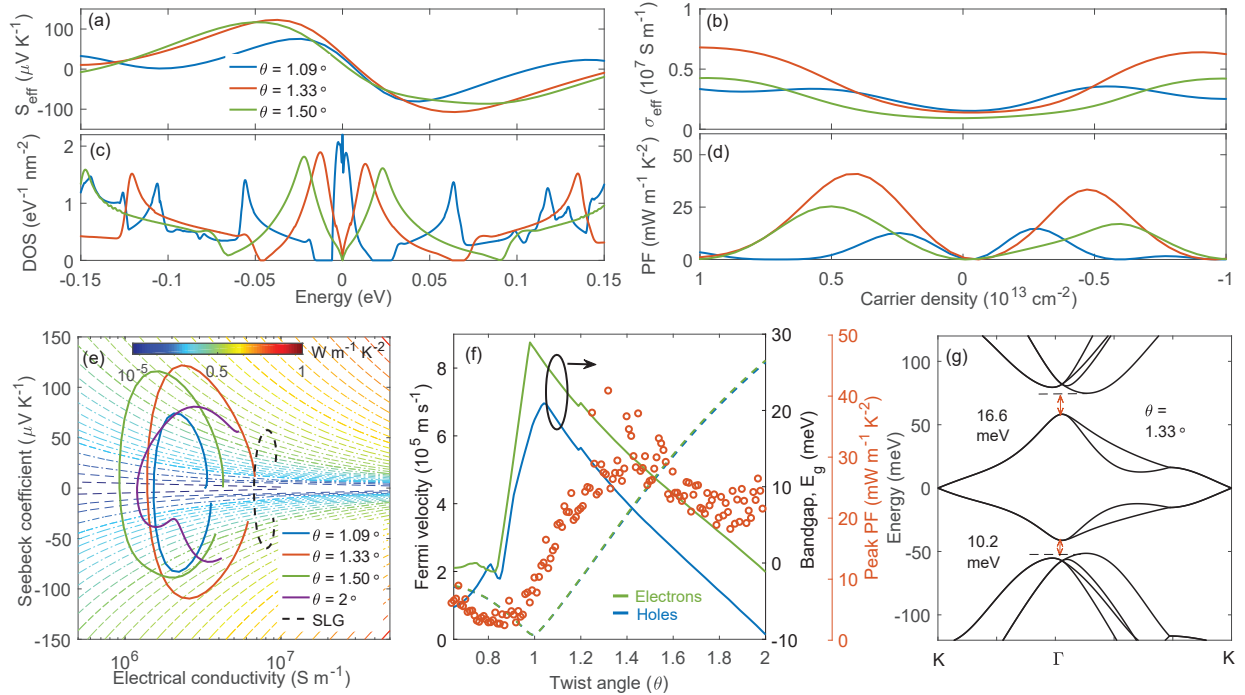}}
	\caption{(a-d) TE properties near magic angle ($\theta$ = 1.09$^\circ$), at twist angle of highest PF ($\theta$ = 1.33$^\circ$), and at twist angle of maximum S$_{eff}$ ($\theta$ = 1.50$^\circ$) a) S$_{eff}$, b) DOS, c) $\sigma_{eff}$, and d) PF at T = 300 K. The secondary gap in the band structure results in a gap in DOS that maximizes the S$_{eff}$ e) Room temperature S$_{eff}$ vs $\sigma_{eff}$ of TBG at different twist angles compared with SLG to understand the PF peak at $\theta$ = 1.28$^\circ$. The constant PF lines are plotted as a guide  to the PF. f) Change in the carrier Fermi velocities of TBG with twist angle. The secondary bandgap (E$_g$) for electrons and holes at different twist angles. g) Band structure, Fermi velocity and band gap of TBG at a twist angle $\theta$ = 1.33$^\circ$ where the highest PF is observed.}
	\label{Fig3}
\end{figure*}

Applying our model, we calculated S$_{eff}$, $\sigma_{eff}$, and PF in the [0 1] crystallographic direction at T = 300 K and plotted them in \ref{Fig2}(c-e) with respect to twist angle and carrier density. Our calculations show that at room temperature, holes have S$_{eff}$ of 122 $\mu$V K$^{-1}$, which is slightly higher than 107 $\mu$V K$^{-1}$ obtained for electrons in Fig. \ref{Fig3}(a). The peak in S$_{eff}$ observed at a twist angle of $\theta$ = 1.33$^\circ$ with a carrier density of 3 $\times$ 10$^{12}$ cm$^{-2}$ in holes and 4 $\times$ 10$^{12}$ cm$^{-2}$ in electrons. The peak in $S_{eff}$ translates to the peak in PF as seen in \ref{Fig2}(c) being significantly higher than SLG. In \ref{Fig3}(a), S$_{eff}$ observed at $\theta$ = 1.33$^\circ$ is compared with S$_{eff}$ near magic angle ($\theta$ = 1.08$^\circ$) and $\theta$ = 1.50$^\circ$. Comparing S$_{eff}$ at these twist-angles with their corresponding DOS (\ref{Fig3}(c)) shows that the peak in S$_{eff}$ is a consequence of DOS band gaps where secondary band gaps (E$_g$) are observed in band structure. The impact of DOS on Seebeck can be qualitatively understood by expanding the Mott formula $S=(\pi^2/3)(k_B^2T/e)\partial \ln\sigma(E)/\partial E|_{E_f}$~\cite{JohnsonPRB}. Approximating the TDF $\sigma(E)$ from Eq.~\ref{TDF} near the Fermi level as $\sigma(E_f)\approx (e/2) v_f(\theta)^2 D(E_f)/S_o$ \cite{HwangPRB09} means that $S\propto D'(E_f)/D(E_f)$, which is maximized when $E_f$ is near a sharp discontinuity in the DOS, in this case a band edge. The gaps in DOS also reduce scattering because inelastic phonon emission from states within an energy $\omega_{op}$ the bottom of a subband is suppressed, which in-turn helps create a narrow TDF, further benefiting thermoelectric properies \cite{KomminiJOPM2019}. 

The simplified expression for TDF has also been widely used to estimate the twist-angle dependence of conductivity from the Fermi velocity as $\sigma(\theta) \approx \sigma(0)[v_f(\theta)/v_f]^2$ \cite{DasSarma2020} at low temperatures and carrier concentrations. However, the TDF also includes the temperature-dependent Fermi window ($\partial f_0/\partial E$) whose width is $\approx 5 k_B T$~\cite{KomminiJOPM2019} so transport occurs over a range of energies around $E_f$, which widens the peak in S$_{eff}$ and the minimum in $\sigma_eff$ as functions of twist angle. 
This trend is illustrated by observing the S$_{eff}$ and the corresponding $\sigma_{eff}$ for different twist angles in \ref{Fig3}(e). Taken together, they produce a peak in PF at room temperature at twist angles around $\theta$ = 1.3$^\circ$ for both electrons and holes, with maximum PF reaching around 40 mWm$^{-1}$ K$^{-2}$ (Fig. \ref{Fig3}(d)). The peaks in PF occur at a carrier density of 3.9 $\times$ 10$^{12}$ cm$^{-2}$ in holes and 4.7 $\times$ 10$^{12}$ cm$^{-2}$ in electrons, near complete band-filling $n_s = \pm$ 4.1 $\times$ 10$^{12}$ cm$^{-2}$. This stands in contrast to other notable properties of TBG, superconductivity and Mott insulating behavior, which typically occur at half-filling. The range of twist angles we simulated domain is limited to angles above 0.65$^\circ$ due to limitations of the continuum model that we used for band structure calculations. This limitation comes from the inclusion of atomic relaxation in TBG that suppress the secondary magic-angle below $\theta$ = 0.65$^\circ$. However, the band gaps occuring at magic angles below 0.65$^\circ$ are smaller \cite{TarnopolskyPRL19}, so we expect them to result in lesser enhancements of Seebeck and PF.

The impact of band gap E$_g$ on S$_{eff}$ is reflected in the PF, as shown by the plot of maximum PF reached either in electrons or holes at each twist angle along with the corresponding electron and hole secondary band gaps (E$_g$) in \ref{Fig3}(f). Both electron and hole band gaps peak near the 1$^\circ$ twist angle due to the gaps and flat bands in the band structure. The flattening is also evident in the corresponding electron and hole Fermi velocities at different twist angles. The growing gap near magic angle is beneficial to S$_{eff}$ as gapping suppresses bipolar transport, while the asymmetry in the DOS increases average carrier energy relative to the Fermi level. On the other hand, lower Fermi velocity is directly reflected in a decrease of conductivity, particularly at low carrier densities. Taken together, the highest PF occurs when the gain in Seebeck overtakes the loss of conductivity, indicated by the peak PF occurring when the E$_g$ and v$_F$ plots cross in \ref{Fig3}(f). 
Fig 3(e) shows the band structure at a twist angle $\theta$ = 1.33$^\circ$, at which the PF peaks, along with the band gap between first and second bands, which is 16.6 meV in electrons and 10.2 meV in holes.


\begin{figure*}[ht!]
	\centering
	\includegraphics[width=1\columnwidth]{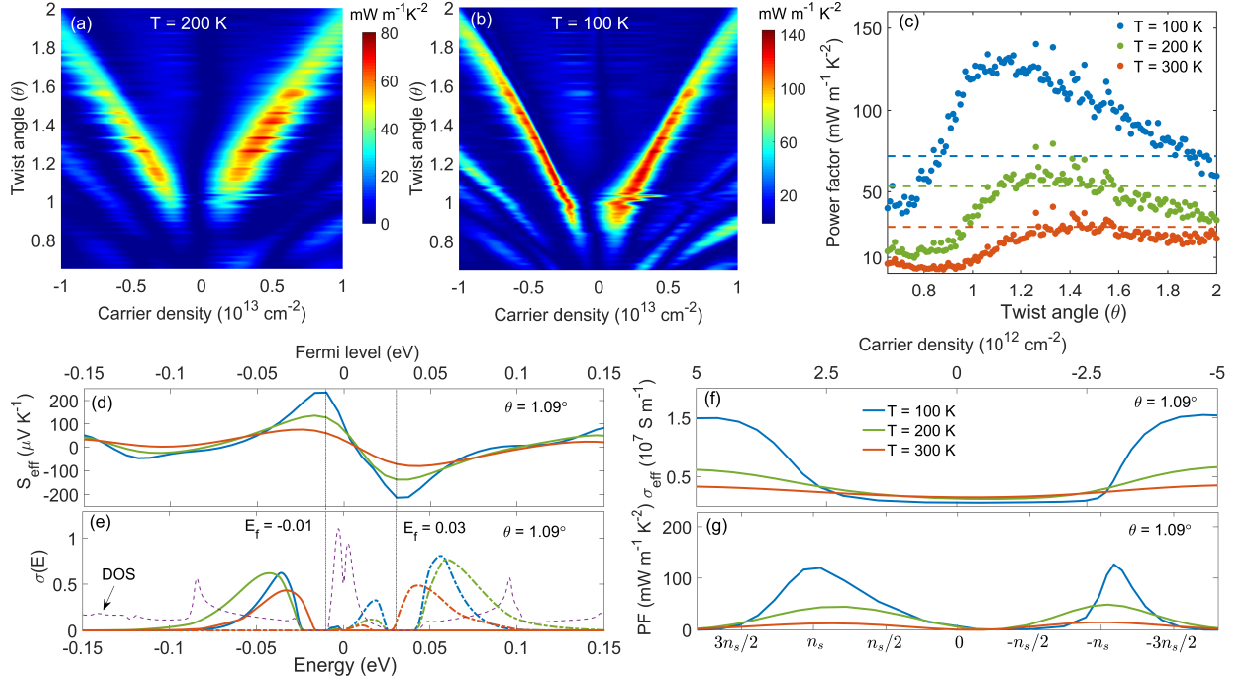}
	\caption{Power factor at a) T = 200 K and b) T = 100 K of TBG at different twist angles. c) Temperature dependence of peak PF at twist angles of TBG, observed by varying the carrier densities. The peak in power factor is observed at $\theta$ = 1.33$^\circ$ at T = 300 K, decreasing slightly to $\theta$ = 1.26$^\circ$ for T = 100 K. d) Temperature dependence of S$_{eff}$ at $\theta$ = 1.09$^\circ$. e) $\sigma(E)$ at Fermi levels where peak in S$_{eff}$ is observed at T = 100 K. S$_{eff}$ peaks when $\sigma(E)$ is asymmetric with respect to the corresponding Fermi level. Variation in f) $\sigma$ and g) power factor with temperature at $\theta$ = 1.09$^\circ$ at different carrier densities. The peak in power factor is observed at a carrier densities near $\pm n_s$ depending on the temperatures.}
	\label{Fig4}
\end{figure*}

To understand the temperature dependence of PF in TBG, we calculated the TE properties at T = 100 K and T = 200 K while changing the twist angle. \ref{Fig4}(a) and \ref{Fig4}(b) shows the PF at different twist angles and carrier densities at T = 100 K and T = 200 K. With decreasing temperature, the carrier concentration and the twist angle at which the PF peaks both decrease. The twist angle at which PF peaks is lowered slightly from $\theta$ = 1.33$^\circ$ at 300 K to $\theta$ = 1.26$^\circ$ at 100 K. The maximum PF either in holes or electrons at T = 100 K and T = 200 K is compared with the values observed at room temperature in Fig.~\ref{Fig4}(c). A maximum PF of 80 mWm$^{-1}$K$^{-2}$ is observed at T = 200 K, increasing to 140 mWm$^{-1}$ K$^{-2}$ at T = 100 K, representing increases of 51 and 100\%, respectively, relative to SLG, which is plotted by the dotted black line in Fig. \ref{Fig4}(c). TBG exhibits increasing power factor with lowered temperature, opposite to the trend observed in SLG \cite{DuanPNAS2016}. This reversal can be explained by the presence of the band gaps in TBG, which help suppress bipolar transport when the gap is larger than the 5 k$_B$T width of the Fermi window $\partial f_0/\partial E$. Similar temperature dependence has been previously observed in high-T$_c$ superconductor FeSe \cite{Shimizu2019}, where the PF in ultrathin samples increased an order in magnitude from 300 K to 100 K. 

The observed peaks in PF coincide with the peak in bandgap between second and first bands for both electrons and holes (shown previously in Fig. \ref{Fig3}(f)). In \ref{Fig4}(d), S$_{eff}$ is plotted near the magic angle at $\theta$ = 1.09$^\circ$, while varying the carrier concentration by moving the position of the Fermi level (E$_f$). The Seebeck coefficient increases with asymmetry in TDF in the vicinity of E$_f$, which is maximized at the edge of a band. Once the lowest subband is completely filled, increasing E$_f$ further causes carriers to fill the second subband. There the DOS flattens, causing $\sigma'(E_f)$ to decrease and eventually change sign, resulting in the reversal of the sign of S$_{eff}$ at high carrier densities seen in FIg.~\ref{Fig4}(d). Contrary to SLG, S$_{eff}$ in TBG increases with decreasing temperature, which can be explained from the TDF $\sigma$(E) shown in \ref{Fig4}(e), where $\sigma$(E) is plotted at the carrier densities where S$_{eff}$ is maximum for holes and electrons, which correspond to $E_f$ = -0.01 and $E_f$ = 0.03, respectively. At lower temperatures, the narrow width of the Fermi window, combined with the band gap, constrains the TDF and maximizes its slope, which increases S$_{eff}$. At this twist angle, $\sigma_{eff}$ is also boosted at lower temperature due to the reduction in linear-in-T carrier-phonon scattering, resulting in a higher PF at low temperatures. Nonetheless, the peak in PF vs. carrier densities remains near $\pm n_s$ at 100 and 200 K, as shown in \ref{Fig4}(g). 

In conclusion, we calculated phonon-limited thermoelectric transport properties of TBG for a range of twist angles, carrier densities, and at several temperatures. Both acoustic and optical phonon scattering rates were included to capture carrier-lattice interactions. Band structure of TBG is obtained from an exact continuum model coupled with tight-binding Hamiltonians, while transport is captured using an iterative solver for the Boltzmann equation. Our calculations show a peak PF of around 40 mWm$^{-1}$ K$^{-2}$ at room temperature, observed at a twist angle of 1.33$^\circ$, which is slightly above the magic angle. The peak PF value is roughly 40\% higher than the highest PF observed in SLG. As twist angle approaches the magic angle, a gap opens in the electronic structure, which improves the Seebeck coefficient, while the Fermi velocity is lowered, depressing conductivity. The gain in the Seebeck outweighs the loss in conductivity, resulting in higher PF at angles slightly above the magic angle. As temperature is lowered, the peak PF increases as the narrower Fermi window helps suppress bipolar transport, reaching 80 and 140 MWm$^{-1}$K$^{-2}$ at 200 and 100 K, respectively.
We conclude that, using temperature and twist angle as knobs, TE properties of TBG can be tuned to achieve superlative performance. Owing to the increase in Seebeck coefficient relative to SLG, junctions between BLG and SLG could form highly sensitive thermocouples. These unique properties of TBG make it a promising candidate for future TE devices that can be operated under a wide spectrum of performance requirements in energy conversion and thermal sensing applications.

\textbf{Acknowledgment.} The authors thank Drs. Efthimios Kaxiras, Jun Yan, and Stephen Carr for fruitful discussions and NSF for financial support through award 1902352.

\bibliography{thermoelectricAK,thermoelectricZA}
\end{document}